\documentclass[11pt]{article}
\usepackage[colorlinks=true, citecolor=blue, urlcolor=blue, linkcolor=blue, breaklinks=true, pdfpagelabels=false]{hyperref}
\usepackage{cite}

\usepackage{hyperref}

\usepackage{amsmath,amsfonts,amssymb}
\usepackage{graphicx}
\usepackage{epsfig,epsf}
\usepackage{bm}
\usepackage{xcolor}

\def\bar {\overline}

\def\be {\begin{equation}}
\def\ee {\end{equation}}
\def\bea {\begin{eqnarray}}
\def\eea {\end{eqnarray}}

\def\barr{\begin{array}}
\def\earr{\end{array}}

\def\opcit(#1){ {\em op. cit.}, #1}

\def\issue(#1,#2,#3){#1, #2 (#3)} 

\def\equationautorefname~#1\null{Eq.\,(#1)\null}
\def\pageautorefname\nobreakspace{p.}

\makeatletter\renewcommand{\p@subsection}{\thesection.}\makeatother%
 
\begin{document}

\renewcommand*{\thefootnote}{\fnsymbol{footnote}}


\begin{center}
{\Large\bf{Updated constraints on triplet vev in the Georgi Machacek model}}


\vspace{5mm}

{\bf Swagata Ghosh} \footnote{swgtghsh54@gmail.com}

\vspace{3mm}
{\em{\ \ \ Department of Physics, Indian Institute of Technology Kharagpur, Kharagpur 721302, India.}}

\end{center}

\begin{abstract}
\noindent
The Georgi-Machacek model containing singly as well as doubly charged Higgs bosons besides more than one neutral Higgs bosons preserves the custodial symmetry at the tree level. 
In addition to the doublet vacuum expectation value (vev) $v_1$ this model has one triplet vev $v_2$, which was most strongly bounded from the above as a function of the custodial triplet mass $m_3$ previously by the $b \rightarrow s \gamma$ data. 
The observed data from the ATLAS and CMS experiments at the LHC for $\sqrt{s} = 13$ TeV for the channels $p p \rightarrow t \bar{t}$, $t \rightarrow H^+ b$, $H^+ \rightarrow c\bar{s}$, $\tau^+ \nu_{\tau}$ for the low mass charged Higgs set new limits on the triplet vev as a function of light charged Higgs boson afterward. 
Subsequent ATLAS data for the decay $H^{\pm} \rightarrow \tau^{\pm} \nu_{\tau}$ updates the limit on $v_2$ from the above as a function of $m_3$. 
The present above limits on $v_2$ is now updated to $6.3 - 13.1$ GeV from the previous value of $7.8 - 17.4$ GeV for the light custodial triplet mass $m_3$ ranging from $130$ to $160$ GeV. 
\end{abstract}



\setcounter{footnote}{0}
\renewcommand*{\thefootnote}{\arabic{footnote}}

\section{Introduction}
\label{intro}

In the year of $2012$ the ATLAS and CMS experiments at the Large Hadron Collider (LHC) reported \cite{ATLAS:2012yve,CMS:2012qbp} the discovery of the Standard Model (SM) Higgs boson with a mass about $125$ GeV. 
Still the experiments at the LHC advance the searches for other non-standard particles, whether they are neutral or charged. 
Together with these exotic particles the $125$ GeV scalar resonance must be comprehended in the scalar sector of any model beyond the Standard Model (BSM). 

The LHC started their searches for the charged scalar and pseudoscalar with mass higher or lower than the top-quark mass $m_t$ since a long time. 
The exploration of light singly charged Higgs bosons includes its decay into $cb$ \cite{Ivina:2022tfm, ATLAS:2021zyv}, $cs$ \cite{CMS:2020osd, ATLAS:2013uxj, ATLAS:2010ofa}, and $\tau\nu_{\tau}$ \cite{CMS:2019bfg, Abbaspour:2018ysj, ATLAS:2016avi, ATLAS:2011pka, ATLAS:2012nhc, Ali:2011qf, CMS:2012fgz, ATLAS:2024hya} at the LHC and other colliders \cite{Hou:2021qff, Akeroyd:2019mvt, Akeroyd:2018axd}. 
The literature also covers the study of the decay of the low mass charged Higgs bosons in different BSM models \cite{Ghosh:2022wbe, Cheung:2022ndq, Akeroyd:2016ymd, Benbrik:2021wyl, Akeroyd:2016ssd, Akeroyd:2012yg, Akeroyd:2022ouy}. 

The Georgi-Machacek (GM) Model \cite{Georgi:1985nv} encloses singly charged as well as doubly charged Higgs bosons in its scalar sector. 
This paper focuses only on the low mass singly charged Higgs boson decay with charged Higgs mass ranging from $80-160$ GeV. 
The indirect constraints coming from the $b \rightarrow s \gamma$ decay \cite{Hartling:2014aga} put the first limits on the triplet vev as a function of the triplet mass in a wide mass range. 
The observed data of ATLAS and CMS for the light charged Higgs bosons decay to $cs$ and $\tau\nu$ in the mass range of $80-160$ GeV put a more stringent bound on the triplet vev \cite{Ghosh:2022wbe} of the GM model for only the low mass range leaving the limit on the triplet vev for the mass range above $160$ GeV unchanged. 
This bound mainly came from the observed data of the ATLAS \cite{ATLAS:2018gfm} for the $H^{\pm} \rightarrow \tau^{\pm}\nu_{\tau}$ decay channel. 
The more recent observed data from ATLAS \cite{ATLAS:2024hya} put stricter upper limit on $v_2$ as a function of low $m_3$. 

The neutral components of the real and the complex triplets in the scalar sector of the Georgi Machacek model acquires the same triplet vev $v_2$. 
Unlike the other scalar triplet extensions of the SM, the GM model protects the tree level $\rho$-parameter, $i.e.$, $M_W^2/M_Z^2 \cos^2 \theta_W = 1$, culminating a substantial value of the $v_2$. 
Two singlets ($h, H$), one triplet ($H_3^0, H_3^+, H_3^-$), and one quintuplet ($H_5^0$, $H_5^+$, $H_5^-$, $H_5^{++}$, $H_5^{--}$) construct the physical scalar sector of the model. 
The physical scalars involve four charge-neutral, four singly-charged and two doubly-charged Higgs bosons. 
The members of the quintuplet do not couple to the SM fermions. 
The couplings of the members of the triplet to the SM fermions are directly proportional to the triplet vev. 

This work is organized as follows. 
Sec. \ref{model} outlines the model along with the related constraints in brief. 
The results of this work are given in Sec. \ref{result}. 
Finally Sec. \ref{conclusions} concludes. 

\section{The Georgi-Machacek Model and the Constraints}\label{model}

In addition to the SM doublet $\left(\phi^{+},\phi^{0}\right)^T$, one $SU(2)_L$ real triplet $\left(\xi^{+},\xi^0,\xi^{-}\right)^T$ and one $SU(2)_L$ complex triplet $\left(\chi^{++},\chi^{+},\chi^0\right)^T$ set up the scalar sector of the Georgi Machacek model. 
The respective hypercharges are $Y=1, 0, 2$. 
The doublet and the triplets form the bi-doublet and bi-triplet respectively \cite{Hartling:2014zca}, as, 
\begin{equation}
\Phi =
\begin{pmatrix}
 \phi^{0*} & \phi^{+}\cr
 \phi^{-} & \phi^0
\end{pmatrix}\,
,\quad
X =
\begin{pmatrix}
 \chi^{0*} & \xi^{+} & \chi^{++}\cr
 \chi^{-} & \xi^0 & \chi^{+}\cr
 \chi^{--} & \xi^{-} & \chi^0
\end{pmatrix}\,.
\end{equation}
The most general scalar potential in terms of $\Phi$ and $X$ is given by,  
\begin{eqnarray}
V\left(\Phi,X\right) &=& \frac{{\mu_2}^2}{2}\, {\rm Tr}\left(\Phi^\dag\Phi\right) 
+ \frac{{\mu_3}^2}{2}\, \rm{Tr}\left(X^\dag X\right)
+ {\lambda_1}\left[{\rm Tr}\left(\Phi^\dag\Phi\right)\right]^2  
+ {\lambda_2}\, {\rm Tr}\left(\Phi^\dag\Phi\right)\, {\rm Tr}\left(X^\dag X\right) \nonumber\\
&& + {\lambda_3}\, {\rm Tr}\left(X^\dag X X^\dag X\right)
 + {\lambda_4}\left[{\rm Tr}\left(X^\dag X\right)\right]^2 
- \frac{{\lambda_5}}{4}\, {\rm Tr}\left(\Phi^\dag \sigma^a \Phi\sigma^b\right)\, {\rm Tr}\left(X^\dag t^a X t^b\right) \nonumber\\
&& - \frac{{M_1}}{4} \, {\rm Tr}\left(\Phi^\dag \sigma^a \Phi\sigma^b\right) {\left(U X U^\dag\right)_{ab}} 
- {M_2} \, {\rm Tr}\left(X^\dag t^a X t^b\right) {\left(U X U^\dag\right)_{ab}}\, ,
\label{eq:genPot} 
\end{eqnarray}
where $\sigma^a$ are the three Pauli matrices. 
The three $t^a$ matrices with $a=1,2,3$ and the matrix $U$ are given by, 
\begin{equation}
t^1=\frac{1}{\sqrt2}
\begin{pmatrix}
 0 & 1 & 0\cr
 1 & 0 & 1\cr
 0 & 1 & 0
\end{pmatrix}
\,,
t^2=\frac{1}{\sqrt2}
\begin{pmatrix}
 0 & -i & 0\cr
 i & 0 & -i\cr
 0 & i & 0
\end{pmatrix}
\,,
t^3=
\begin{pmatrix}
 1 & 0 & 0\cr
 0 & 0 & 0\cr
 0 & 0 & -1
\end{pmatrix}\,,
U=
\frac{1}{\sqrt{2}}\begin{pmatrix}
 -1 & 0 & 1\cr
 -i & 0 & -i \cr
 0 & \sqrt{2} & 0
\end{pmatrix}\,.
\end{equation}
The doublet vev $v_1$ and the triplet vev $v_2$, which are acquired after the electroweak symmetry breaking (EWSB), are related to the electroweak vev $v$ as $\sqrt{v_1^2 + 8 v_2^2} = v \approx 246$ GeV. 
The preservation of the tree level $\rho$-parameter to unity is a direct consequence of the equality of the real and the complex triplet vevs. 

The scalar sector of the GM model consists of a custodial quintuplet $H_5$ ($H_5^{++}, H_5^{+}, H_5^0, H_5^{-}, H_5^{--}$), one custodial triplet $H_3$ ($H_3^{+},H_3^0,H_3^{-}$), and two singlets ($h, H$). 
The expressions of these ten physical fields in terms of the component fields and the vevs are given by,
\begin{eqnarray}
&&H_5^{\pm\pm}=\chi^{\pm\pm}\,, \ \ \ 
H_5^{\pm}=\frac{\left(\chi^{\pm}-\xi^{\pm}\right)}{\sqrt{2}}\,, \ \ \ 
H_5^0=\sqrt{\frac23}\xi^0-\sqrt{\frac13}\chi^{0R}\,, \nonumber\\
&&H_3^{\pm}=-2\sqrt{2}\,\frac{v_2}{v}\, \phi^{\pm}+ \frac{v_1}{v}\, \frac{\left(\chi^{\pm}+\xi^{\pm}\right)}{\sqrt{2}}\,, \ \ \ 
H_3^0=-2\sqrt{2}\,\frac{v_2}{v}\, \phi^{0I}+\frac{v_1}{v}\, \chi^{0I}\,, \nonumber\\
&&h = \cos {\alpha}\,\, \phi^{0R}-\sin {\alpha}\,\, \left(\sqrt{\frac13}\xi^0+\sqrt{\frac23}\chi^{0R}\,\right)\,, \nonumber\\ 
&&H = \sin {\alpha}\,\, \phi^{0R}+\cos {\alpha}\,\, \left(\sqrt{\frac13}\xi^0+\sqrt{\frac23}\chi^{0R}\,\right)\,.
\label{eq:fields}
\end{eqnarray}
The neutral scalar mixing angle $\alpha$ can be expressed in terms of the potential parameters and the vevs as,
\begin{equation}
\tan{2\alpha}=\frac{4\sqrt{3} v_1 v_2 \left[-M_1+4\left(2\lambda_2-\lambda_5\right)v_2\right]}{ M_1 v_1^2 - 24 M_2 v_2^2 + 32 (\lambda_3 + 3 \lambda_4) v_2^3 - 32 {\lambda_1}{v_1}^2 v_2 }\,.
\end{equation}
The mass of $H$ is greater than the mass of $h$. 
The members of $H_3$ and $H_5$ share the degenerate mass $m_3$ and $m_5$ respectively. 
In terms of the quartic parameters ($\lambda_{1-5}$) and the trilinear parameters ($M_{1,2}$) present in the potential together with the vevs, the mass-squared of the physical scalars are given by, 
\newpage
\begin{eqnarray}
{m_{h,H}}^2&=&\frac12 \Bigg[8 {\lambda_1}{v_1}^2 + \frac{M_1 v_1^2}{4 v_2} - 6 M_2 v_2 + 8 (\lambda_3 + 3 \lambda_4) v_2^2\, \nonumber\\ 
&&\mp\sqrt{\left(8 {\lambda_1}{v_1}^2 - \frac{M_1 v_1^2}{4 v_2} + 6 M_2 v_2 - 8 (\lambda_3 + 3 \lambda_4) v_2^2
 \right)^2 + 3 v_1^2 \left(-M_1+4\left(2\lambda_2-\lambda_5\right)\right)^2}\Bigg]\,, \nonumber\\
 {m_3}^2&=&\left(\frac{M_1}{4{v_2}}+\frac{\lambda_5}{2}\right)v^2\,, \nonumber\\
 {m_5}^2&=&\frac{M_1}{4{v_2}}{v_1}^2+12{M_2}{v_2}+\frac32{\lambda_5}{v_1}^2+8{\lambda_3}{v_2}^2\,.
\label{eq:masses}
\end{eqnarray}
This work considers $m_h \approx 125$ GeV such that the lighter physical singlet $h$ given in the Eqn. (\ref{eq:fields}) is the SM Higgs boson. 

The square of the bilinear coefficients ($\mu_{2,3}$) of the potential may be expressed in terms of the vevs as well as the trilinear coefficients ($M_{1,2}$) and the quartic coefficients ($\lambda_{1-5}$) present in the potential as,
\begin{eqnarray}
{\mu_2}^2&=&-4{\lambda_1}{v_1}^2-3\left(2{\lambda_2}-{\lambda_5}\right){v_2}^2
+\frac32{M_1}{v_2} \,,\nonumber\\
{\mu_3}^2&=&-\left(2{\lambda_2}-{\lambda_5}\right){v_1}^2-4\left({\lambda_3}+3{\lambda_4}\right){v_2}^2
+\frac{M_1 {v_1}^2}{4 v_2}+6{M_2}{v_2} \,,
\label{eq:mu}
\end{eqnarray}
where the quartic coefficients can be expressed in terms of the four physical masses, $m_h$, $m_H$, $m_3$, $m_5$, and the mixing angle $\alpha$ as,
\begin{eqnarray}
\lambda_1 &=& \frac{1}{8 v_1^2}\left(m_h^2\,\cos^2{\alpha} + m_H^2\,\sin^2{\alpha}\right)\,, \nonumber\\
\lambda_2 &=& 
\frac{1}{8 v_2}\left[\left(m_H^2-m_h^2\right)\,\frac{1}{\sqrt{3}v_1}\sin{2\alpha}
-M_1 +8 m_3^2 \frac{v_2}{v^2} \right]\,, \nonumber\\
\lambda_3 &=& \frac{1}{8 v_2^2} \left( m_5^2 -3 m_3^2 \frac{v_1^2}{v^2} 
+\frac{M_1 v_1^2}{2 v_2} - 12 M_2 v_2 \right)\,, \nonumber\\
\lambda_4 &=& \frac{1}{24 v_2^2} \left( m_h^2\,\sin^2 {\alpha} + m_H^2\,\cos^2 {\alpha} - m_5^2
+ 3 m_3^2 \frac{v_1^2}{v^2} - 3 M_1 \frac{v_1^2}{4 v_2} + 18 M_2 v_2 \right)\,, \nonumber\\
\lambda_5 &=& 2 \left( \frac{m_3^2}{v^2} - \frac{M_1}{4 v_2} \right)\,.
\label{eq:lamda}
\end{eqnarray}
These quartic coefficients are heavily constrained by the perturbative unitarity and electroweak vacuum stability following \cite{Hartling:2014zca, Hartling:2014aga}. 
Constraints coming from the perturbative unitarity are given by,
\begin{eqnarray}
\sqrt{\left( 6\lambda_1-7\lambda_3-11\lambda_4 \right)^2+36\lambda_2^2}+\mid 6\lambda_1+7\lambda_3+11\lambda_4\mid &<& 4\pi\,, \nonumber\\
\sqrt{\left( 2\lambda_1+\lambda_3-2\lambda_4 \right)^2+\lambda_5^2}+\mid 2\lambda_1-\lambda_3+2\lambda_4\mid &<& 4\pi\,, \nonumber\\
\mid 2\lambda_3+\lambda_4\mid &<& \pi\,, \nonumber\\
\mid \lambda_2-\lambda_5\mid &<& 2\pi\,,
\label{eq:unitarity}
\end{eqnarray}
and the constraints coming from the electroweak vacuum stability are given by,
\begin{eqnarray}
\lambda_1\, &>&\, 0\,,\nonumber\\
\lambda_2+\lambda_3\,&>&\,0\,,\nonumber\\
\lambda_2+\frac12\lambda_3\,&>&\,0\,,\nonumber\\
-\mid\lambda_4\mid +2\sqrt{\lambda_1\left(\lambda_2+\lambda_3\right)}\,&>&\,0\,,\nonumber\\
\lambda_4-\frac14 \mid\lambda_5\mid +\sqrt{2\lambda_1\left(2\lambda_2+\lambda_3\right)}\,&>&\,0\,.
\label{eq:stability}
\end{eqnarray}
Also the LHC Higgs signal strength data \cite{CMS:2018lkl,ATLAS:2019slw} obtained from the ATLAS and CMS experiments at $\sqrt{s}=13$ TeV curb the parameter space of the Georgi Machacek model. 
Since $h$ is considered as the SM Higgs boson, one must know how this $h$ couples to the SM fermions $f$ and vector bosons $V$. 
For this purpose, the ratio of couplings of $h$ with $f$ and $V$ in the GM model to that in the SM are required and they are given by, 
\begin{equation}
\kappa_f^h = \frac{v}{v_1}c_{\alpha}\,,\quad
\kappa_V^h = -\frac{1}{3v}(8\sqrt{3}s_{\alpha}v_2\,-\,3 c_{\alpha}v_1)\,.
\label{eq:kappa}
\end{equation}
The charged Higgs bosons $H_{3,5}^{\pm},\, H_5^{\pm\pm}$ contribute at the loop level for the $h \rightarrow \gamma \gamma$ decay in this model, and hence the corresponding couplings of these charged particles with the SM Higgs boson are listed below :
\begin{eqnarray}
-i g_{h{H_3^+}{H_3^-}} &=&
-i (64\lambda_1 c_{\alpha}\frac{v_2^2 v_1}{v^2}\,-\frac{8}{\sqrt{3}}
\frac{v_1^2 v_2}{v^2}s_{\alpha}(\lambda_3+3\lambda_4)\,- \frac{4}{\sqrt{3}} \frac{v_2 M_1}{v^2} (s_{\alpha}v_2-\sqrt{3}c_{\alpha}v_1)\, \nonumber\\
&&-\frac{16}{\sqrt{3}}\frac{v_2^3}{v^2}s_{\alpha} (6\lambda_2+\lambda_5)\, -\,c_{\alpha}\frac{v_1^3}{v^2}(\lambda_5-4\lambda_2)\,+\,
2\sqrt{3}M_2\frac{v_1^2}{v^2}s_{\alpha}\, \nonumber\\
&& -\,\frac{8}{\sqrt{3}}\lambda_5\frac{v_1 v_2}{v^2} (s_{\alpha}v_1\,-\,\sqrt{3}c_{\alpha}v_2))\,, \nonumber \\
-i g_{h{H_5^+}{H_5^-}} &=& -i g_{h{H_5^{++}}{H_5^{--}}} = 
-i( -8\sqrt{3}(\lambda_3+\lambda_4) s_{\alpha} v_2\,+\,(4\lambda_2+\lambda_5)c_{\alpha}v_1
-\,2\sqrt{3}M_2 s_{\alpha} )\,.\nonumber\\
\label{eq:chargedcoupling}
\end{eqnarray}
Since only the decay of the charged Higgs boson $H_3^{\pm}$ to $\tau\nu_{\tau}$ is the interest of this work, here we enlist the corresponding couplings 
\begin{eqnarray}
 H_3^+\overline{\nu}l : i\,\, 4\,\,\frac{v_2}{v_1}\,\, \frac{m_l}{v}P_R \,,
 \quad
 H_3^-\overline{l}\nu : i\,\, 4\,\,\frac{v_2}{v_1}\,\, \frac{m_l}{v}P_L\,,
\label{eq:H3pmtaunu}
\end{eqnarray}
for the convenience of the reader. 
Here $P_{R,L} = (1 \pm \gamma_5)/2$ are the projection operators. 

\section{Results}\label{result}

For the production and decay of the singly charged scalars in the low mass region, with the scalar mass below the top-quark mass, the production channel considered is $pp \rightarrow t\overline{t}$, $t \rightarrow H^+b$. 
The cross-section of this production mechanism is greater than that of the Drell-Yan process \cite{Ghosh:2022wbe}. 
The Georgi Machacek model contains four singly charged scalars $H_{3,5}^{\pm}$ out of which only two ($H_3^{\pm}$) couple with the fermions. 
Therefore, the singly charged Higgs we are considering here is $H_3^{\pm}$. 
Regarding the decay process of the light $H_3^{\pm}$, following \cite{Ghosh:2022wbe} one can see that, though the branching ratio of $H_3^{\pm} \rightarrow cs$ is higher than that of $H_3^{\pm} \rightarrow \tau^{\pm}\nu_{\tau}$, the experimental result mainly the ATLAS data for the latter is of our interest to update the upper bound on the triplet vev $v_2$ as a function of the triplet mass $m_3$. 
The bound on $v_2$ as a function of $m_3$ is also obtained from ATLAS and CMS data for $H_3^{\pm} \rightarrow cs$ and the CMS data for $H_3^{\pm} \rightarrow \tau^{\pm} \nu_{\tau}$, but they do not provide the bound as stricter as provided by the $H_3^{\pm} \rightarrow \tau^{\pm}\nu_{\tau}$ ATLAS data. 

\begin{figure}
 \begin{center}
 \includegraphics[width= 6cm]{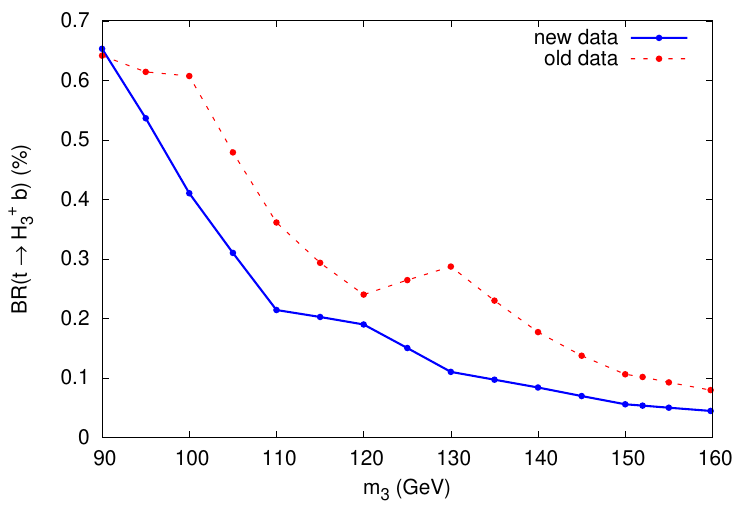} \ \ 
  \includegraphics[width= 6cm]{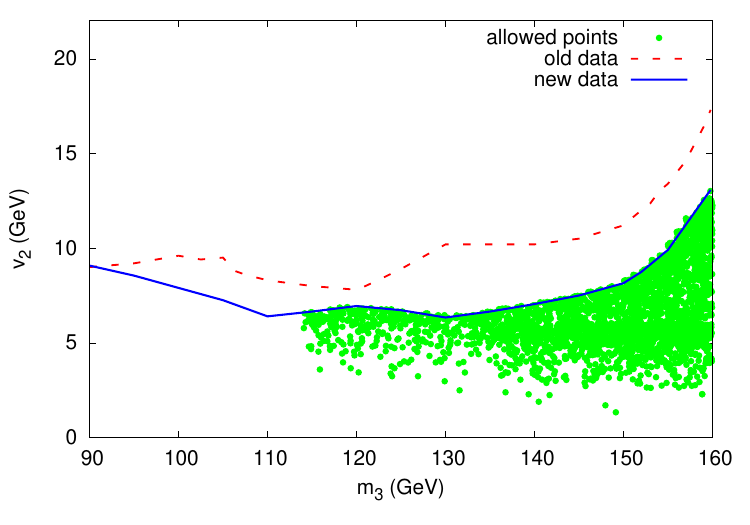} \\
 \end{center}
 \caption{\small The upper limit on the (Left) Branching ratio of $t \rightarrow H_3^+ b$ in percentage and (Right) triplet vev $v_2$ as a function of the triplet mass $m_3$.
 The dashed red line and the solid blue line correspond to the old \cite{ATLAS:2018gfm, Ghosh:2022wbe} and the new \cite{ATLAS:2024hya} ATLAS data for the decay $H^{\pm} \rightarrow \tau^{\pm} \nu_{\tau}$ respectively. 
 The green points in the $m_3 - v_2$ plane are allowed by the theoretical constraints of the GM model, LHC Higgs signal strength data at $\sqrt{s}=13$ TeV, all the ATLAS and CMS data for the fermionic decay of the low mass singly charged scalar.} 
 \label{fig:figure}
 \end{figure}

After the previous data \cite{ATLAS:2018gfm}, ATLAS reported a new data \cite{ATLAS:2024hya} which make us to investigate the channel $pp \rightarrow t\overline{t}$, $t \rightarrow H_3^+ b$, $H_3^+ \rightarrow \tau^+ \nu_{\tau}$ in the context of the GM model. 
The careful study of this channel along with the theoretical constraints of the model and the LHC Higgs signal strength data at $\sqrt{s} = 13$ TeV show that the upper bound on $v_2$ can be updated from the previous bound obtained in the reference \cite{Ghosh:2022wbe}. 
For this, we scan the model parameters in the range as follows :
\begin{eqnarray}
130\, \text{GeV} \le m_H \le 300\, \text{GeV}\,,\,\,
90\, \text{GeV} \le m_3 \le 160\, \text{GeV}\,,\,\,
80\, \text{GeV} \le m_5 \le 250\, \text{GeV}\,,\,\,\nonumber\\
-1000\, \text{GeV} \le M_{1,2} \le 1000\, \text{GeV}\,,\,\,
-1 \le \sin \alpha \le 1\,,\,\,
1\, \text{GeV} \le v_2 \le 18\, \text{GeV}\,.\nonumber
\end{eqnarray}
We consider $h$ as the SM Higgs and set $m_h = 125$ GeV. 
To obtain the limit on the parameter space provided by the ATLAS data we used FeynRules \cite{Alloul:2013bka} and MadGraph5 (v2.8.2) \cite{Alwall:2014hca}. 
The allowed parameter space in the $m_3 - v_2$ plane are depicted by the green points in the Fig. \ref{fig:figure}. 
Next we set 
\begin{equation}
m_H = 200\,~\text{GeV}\,,\,\,m_5 = 85\,~\text{GeV}\,,\,\,
\sin{\alpha} = -0.2\,,\,\,M_1 = M_2 = 100\,~\text{GeV}\,,\,\,\nonumber
\end{equation}
and vary $m_3$ and $v_2$ in the range of $90 - 160$ GeV and $2 - 18$ GeV respectively to get the corresponding upper bound on the triplet vev as a function of the triplet mass as obtained from the newly reported ATLAS data \cite{ATLAS:2024hya}. 
The solid blue line in the $m_3 - v_2$ plane depicts this upper bound. 
One can easily find that this bound is more stringent than the bound represented by the dashed red line that was obtained previously \cite{Ghosh:2022wbe} from the older ATLAS data \cite{ATLAS:2018gfm}. 
The upper limit on $v_2$ is updated to $6.33 - 13.1$ GeV from $7.8 - 17.4$ GeV. 
One can also note that the corresponding triplet mass for the minimum value of the triplet vev is now shifted to $m_3 \approx 130$ GeV from $m_3 \approx 120$ GeV, though the maximum value of $v_2$ is still at $m_3 \approx 160$ GeV. 
The left plot of the Fig. \ref{fig:figure} also presents the corresponding upper limit on the branching ratio of $t \rightarrow H_3^+b$ as a function of $m_3$. 
Like the old limit in the dashed red line, the branching ratio corresponding to the new data given in the solid blue line is still below $0.7\%$, though it is much lowered now. 

\section{Conclusions}\label{conclusions}

The upper limit on the triplet vev $v_2$ as a function of the custodial triplet mass $m_3$ ranging from $10 - 1000$ GeV was previously given by the indirect constraints. 
The most strict bound came from the $b \rightarrow s \gamma$ decay. 
This was valid for the Georgi Machacek model, which is a triplet extension model of the SM. 
The scalar sector of this model contains one $SU(2)_L$ real triplet and one $SU(2)_L$ complex triplet in addition to the SM complex doublet such that the tree level $\rho$-parameter is conserved to give equal triplet vev. 
Besides the neutral scalars ($h,\, H,\, H_3^0,\, H_5^0$), the GM model also consists of two doubly charged scalars ($H_5^{\pm\pm}$) and four singly charged Higgs bosons ($H_3^{\pm},\, H_5^{\pm}$). 

The recently reported data of the LHC, both ATLAS and CMS, for the decay channels $H^{\pm} \rightarrow cs\,\, \rm{or}\,\, \tau^{\pm}\nu_{\tau}$ at $\sqrt{s} = 13$ TeV, specially at the low mass region where the mass of the charged scalar is below the mass of the top-quark, encourage us for the investigation of these decay channels in the context of the Georgi Machacek model. 
The singly charged scalars $H_5^{\pm}$ of the GM model do not couple to the fermions and hence we are interested in the fermionic decay of the low mass $H_3^{\pm}$ with $m_3 \le m_t$. 
The production channel we considered here following the LHC searches is $t(\overline{t}) \rightarrow H_3^{+(-)}b$, where the corresponding branching ratio is always less than $1\%$. 
The previous study in the context of the Georgi Machacek model considering $t \rightarrow H_3^+ b$, $H_3^+ \rightarrow c\overline{s}$ or $\tau^+ \nu_{\tau}$ data from ATLAS and CMS separately, together with the theoretical constraints and the LHC Higgs signal strength data, showed that when $m_3$ is in between $90 - 160$ GeV, the upper limit of the triplet vev $v_2$ as a function of $m_3$ is much lower than the limit obtained from $b \rightarrow s \gamma$ data. 
The most stringent bound came from the ATLAS data for the decay channel $H_3^{\pm} \rightarrow \tau^{\pm}\nu_{\tau}$, as the upper limit on $v_2$ was nearly in between $7.8 - 17.4$ GeV. 
Most recently, ATLAS again reported a new set of data for the same decay channel and this result is followed up in this very work to give a new updated upper bound on the triplet vev $v_2$ approximately in the range of $6.33 - 13.1$ GeV, which is much lower than the previous bound as a function of the triplet mass $m_3$ in between $90 - 160$ GeV. 
The triplet mass for the lowest upper bound on $v_2$ was previously at about $m_3 \approx 120$ GeV, but now at $m_3 \approx 130$ GeV. 
Though with more updates in the experimental data this limit can still be updated, this is the latest updated limit on the triplet vev in the Georgi Machacek model till date. 

\vspace{0.5cm}
{\em{\bf Acknowledgements}} --- The author would like to acknowledge the Government of India for financial support through ANRF-NPDF scholarship with grant no. PDF/2022/001784.


\end{document}